\documentclass[onecolumn,authoryear]{els-mrw} 

\usepackage{amsmath,amssymb,amsfonts,amsthm,makeidx,graphicx}
\usepackage{txfonts}
\usepackage{helvet}
\usepackage{natbib}
\usepackage[svgnames]{xcolor}

\def\aj{AJ}%
\def\apj{ApJ}%
\def\apjl{ApJL}%
\def\apjs{ApJS}%
\def\apss{ApSS}%
\def\aap{A\&A}%
\def\araa{ARA\&A}
\def\mnras{MNRAS}%
\def\nar{New~Astron.~Rev.}
\def\pasa{PASA}%
\def\pasj{PASJ}%
\def\zap{Zeitschrift f\"ur Astrophysik}

\def\msol{\ensuremath{M_\odot}}

\def\teff{\ensuremath{T_\text{eff}}}
\def\hii{H{\sc ii}}
\def\tria{\ensuremath{3\alpha}}
\def\isotop#1#2{\ensuremath{^{#1}\text{#2}}}
\def\cago{\ensuremath{\isotop{12}{C}(\alpha,\gamma)\isotop{16}{O}}}

\def\encyclochap#1{Chapter on "{\textbf{\color{Blue}#1}}" in this Encyclopedia}

\begin{document}

\chapter{Evolution and final fates of massive stars}\label{chap1}

\author[1]{Sylvia Ekstr\"om}%


\address[1]{\orgname{University of Geneva}, \orgdiv{Department of astronomy}, \orgaddress{Chemin Pegasi 51, 1290 Versoix GE, Switzerland}}

\articletag{Chapter Article tagline: update of previous edition,, reprint..}

\maketitle


\begin{glossary}[Nomenclature]
\begin{tabular}{@{}lp{34pc}@{}}
BH & Black Hole\\
CBM & Convective-Boundary Mixing\\
CSM & Circumstellar Medium\\
HRD & Hertzsprung-Russell diagram\\
IMF &Initial Mass Function\\
ISM & Interstellar Medium\\
LBV & Luminous Blue Variable\\
MLT & Mixing Length Theory\\
MS &Main Sequence\\
NS & Neutron Star\\
RSG & Red Supergiant\\
SN & Supernova\\
TAMS &Terminal-Age Main Sequence\\
WR & Wolf-Rayet stars\\
ZAMS &Zero-Age Main Sequence\\
\end{tabular}
\end{glossary}

\begin{abstract}[Abstract]
Massive stars are able to pursue their evolution through the whole sequence of burning phases. They are born hot and luminous, and live a short life before exploding as a supernova or collapsing directly into a black hole. They have a strong impact on their surrounding, injecting mechanical energy, ionising radiation, and nucleosynthetic products in the interstellar medium. They are the driver of galaxy evolution and trigger star formation. Their high luminosity makes them visible in distant galaxies, and some of them are standard candles we use to root the distance ladder of the Universe. This chapter describes the status of our knowledge about massive stars and the nucleosynthetic path they go through the different phases of their evolution.
\end{abstract}

\noindent{\large{\bf Keywords:}} Stellar Physics - Stellar evolution - Stellar structures - Stellar nucleosynthesis

\section*{Keypoints}
\begin{itemize}
  \item Massive stars go through all the nuclear burning phases (H - He - C - Ne - O - Si).
  \item They are powerful sources of kinetic energy and radiation in galaxies.
  \item The are key players in the chemical evolution of the Universe.
  \item Some of them are used as a crucial step in the building of the cosmological distance ladder.
\end{itemize}

\section{Introduction\label{sec:intro}}
Massive stars are defined as stars that are massive enough to be able to evolve beyond helium burning into the advanced phases of nuclear burning (C, Ne. O, Si). Depending on the physics considered, the transition between intermediate-mass stars and massive stars is around 8-10 {\msol}. Below this threshold, their core is unable to reach the high temperature needed to ignite carbon burning. On the high mass end, we usually define as very massive stars (VMS) the stars that have an initial mass above $\sim$100 and up to 300-500\,{\msol}, and as super massive stars (SMS) the ones that start their nuclear life with more than 500\,{\msol}, up to the order of 10$^4$\,\msol. While some VMS are observed in low metallicity environments \citep{Crowther2010a}, SMS are purely theoretical objects as of today. In what follows, the term {\it massive stars} will be used globally for all stars above 8\,\msol.

\begin{figure}[t]
\centering
\includegraphics[width=.75\textwidth]{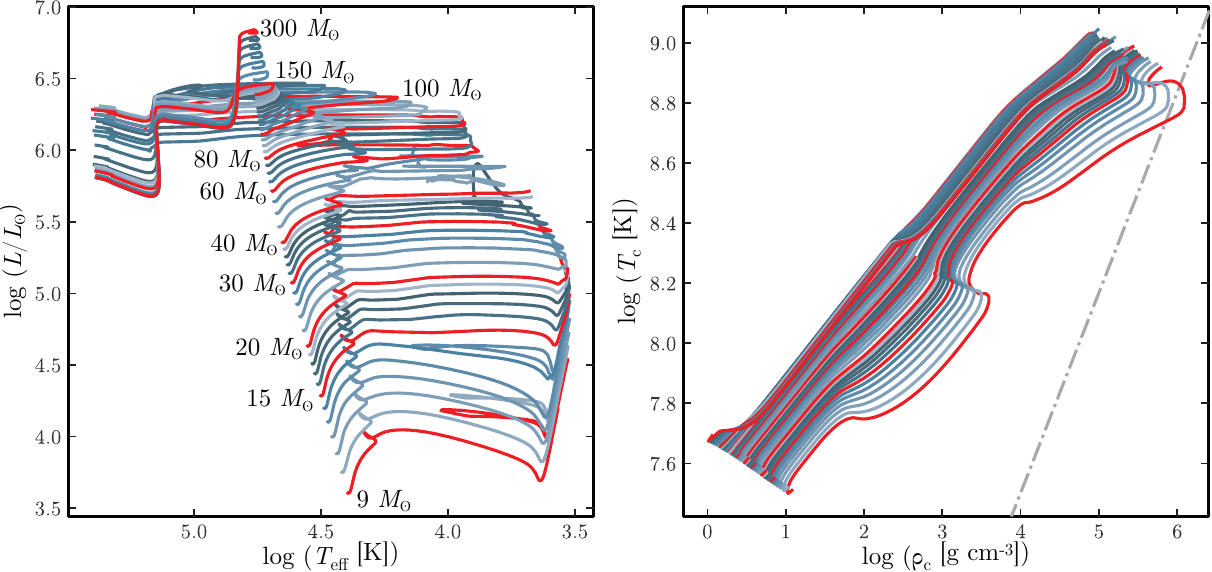}
\caption{Non rotating massive models from the MIST library \citep{Choi2016a}, at solar metallicity. Masses are ranging from 9 to 300\,\msol, with a few ones labelled and drawn in red. {\it Left:} Surface conditions on a Hertzsprung-Russell diagram; {\it Right:} Central conditions on a $\log(T_\text{c})$ vs $\log(\rho_\text{c})$ diagram. The dash-dotted line marks the transition between perfect gas and degenerate conditions.}
\label{fig:mist_hrd}
\end{figure}

Massive stars occupy the higher end of the Hertzsprung-Russell diagram (HRD), with a luminosity ranging from 4000 to millions of times the solar one (see Fig.~\ref{fig:mist_hrd}).
This makes them far-seen beacons in distant galaxies, and an important step in the distance ladder of the Universe \citep{Kudritzki2012a}.

Compared to low-mass stars, a massive star lifetime is very short: a few millions or tens of million years only. Although they benefit from a large reservoir, they burn their fuel at a faster pace because their core is extremely hot. Schematically, the luminosity is proportional to the mass as $L\propto M^3$. A simple estimation of the lifetime of a star is the energy reservoir divided by the luminosity: $\tau\propto\frac{E}{L}$. Taking $E$ as $E\sim fMc^2$ (where $f$ is the fraction of the star's mass participating in the nuclear burning), and with the $M-L$ relation, this yields to $\tau\propto M^{-2}$. This brevity makes them tracers of star-forming regions, because they don't have time to leave their birth place before dying in a supernova explosion or a direct collapse into a black hole.

Massive stars are rare objects: only 2-3 stars more massive than 8\,{\msol} are formed for 1000 stars of low- or intermediate-mass. However their imprint is large. They are responsible for the ionisation of star-forming regions through their UV radiation. Their winds shape the inter-stellar medium (ISM). They are the first actors of chemical evolution in the early Universe, and are the sources of some elements that are crucial for life, like oxygen or phosphorus \citep{Kobayashi2020a,Masseron2020a}. Because of their rarity, until recently only a few hundreds of massive stars were observed, and a large part of our knowledge about those extreme objects relied on numerical modelling. Recent or coming large surveys (such as IACOB, VFTS, SDSS-V, 4MOST, Gaia, LSST, among others) are progressively changing the situation, increasing by orders of magnitude the number of observed massive stars. Our knowledge on the stellar physics at play in massive stars will soon get to be rooted in statistically significant observational samples, but for now, we are still left with a  knowledge flawed by a lot of uncertainties and variations of physical ingredients from code to code.

\section{Modelling massive stars: ingredients and uncertainties\label{sec:intro:uncertain}}
Numerical models show that massive stars evolution is strongly affected by various processes, among which the dominant ones are the mass loss\footnote{see also the \encyclochap{Stellar winds})}and the internal transport processes, for both the angular momentum and the chemicals\footnote{see also the \encyclochap{Mixing processes in stars}}. Unfortunately, these processes are largely unconstrained as of today.

\subsection{Mass loss}
The strong radiation of massive stars is able to trigger a high mass loss through radiatively-driven winds \citep{Crowther2001a}. In the upper end of the mass domain, stars are very close to the Eddington limit, which can increase the mass loss even more \citep{Bestenlehner2020a}. Losing mass uncovers deeper and deeper layers, until the layers that have been transformed by nuclear reactions (or by internal mixing propagating the modified composition of the core) are exposed at the surface. The subsequent change of gravity and of opacity influences the colour and luminosity of the star, those two in turn having a feedback on the mass-loss rates.

There are still large uncertainties in the mass-loss rates of main sequence (MS) massive stars, and many recipes exist, either empirical or theoretical, to be implemented in stellar evolution codes. In the last decades, observational evidences of clumping in the wind \citep{Bouret2005a,Fullerton2006a} have led to a general downward revision of the rates, with many consequences for stellar populations, like the lower probability of a single star to become a Wolf-Rayet star (WR), or the larger fraction of rapidly-rotating stars predicted \citep{Hirschi2008a}. Even more complicated, in the advanced phases of the evolution, the driving of the wind changes \citep{Humphreys2022a} and eruptive mass loss can occur, as for example in the Luminous Blue Variables phase (LBV) or in the Red Supergiant phase (RSG). By definition, eruptions occur on a short timespan, and with an amplitude that is most probably impossible to predict. However they should be added to the mass loss budget in order for models to capture the full mass evolution.

Mass loss is an important ingredient of stellar modelling, since it removes not only mass but also angular momentum, influencing the evolutionary path as well as the mass and angular momentum of the end point (neutron star or black hole).

\subsection{Convection}
Another complicated ingredient we need to incorporate in the models is convection. Convection being a highly multi-dimension turbulent process, we rely on prescriptions to implement it in the 1D codes used to compute the secular evolution of stars. First the radiative or convective nature of the matter has to be defined through a criterium that can be the Schwarzschild or the Ledoux one \citep{Schwarzschild1958a,Ledoux1947a}. The Schwarzschild criterium is defined as
\begin{equation}
\nabla_\text{rad}>\nabla_\text{ad}
\label{Eq:Schw}
\end{equation}
with the adiabatic gradient $\nabla_\text{ad}=\left(\frac{\partial\ln T}{\partial\ln P}\right)_\text{ad}=\frac{P\delta}{C_P\rho T}$ (where $\delta$ comes from a general equation of state of the form $\Delta\ln\rho = \alpha\,\Delta\ln P - \delta\,\Delta\ln T + \varphi\,\Delta\ln\mu$), and the radiative gradient $\nabla_\text{rad}=\frac{3}{16\pi a c G}\frac{\kappa LP}{MT^4}$ (where $\kappa$ the opacity and obvious meanings for $L$, $P$, $M$, and $T$). The Ledoux criterium includes the restoring effect of chemical gradients:
\begin{equation}
\nabla_\text{rad}>\nabla_\text{ad}+\frac{\varphi}{\delta}\nabla_\mu
\end{equation}
with $\mu$ the average mean molecular weight. From observations, it is not clear which criterium should be adopted.

While the acceleration of a convective cell stops when the conditions for convection are not fulfilled anymore, the velocity of the cell does not drop to zero immediately. To account for the incursion of convective cells into the radiatively-stratified zone \citep{Arnett2019a}, the concept of convective-boundary mixing (CBM) was introduced in stellar models. Several implementations exist: step overshoot, where the border of the convective core is displaced by a distance $d_\text{ov}$ that is a fraction of the pressure scale height: $d_\text{ov}=\alpha_\text{ov}H_P$ \citep{Shaviv1973a}; diffusive overshoot, where a decaying diffusive coefficient $D_\text{ov}$ is applied at the border of the core: $D_\text{ov}=D_0 \exp\left(2\frac{r-r_0}{f_\text{ov}H_P}\right)$, with $D_0$ usually taken as a function of the velocity coming from the mixing-length theory \citep[MLT][]{BoehmVitense1958a} $D_0 = V_\text{MLT}\,H_P$ \citep{Herwig1997a}; entrainment, where the border of the convective core progresses outwards in time \citep{Staritsin2013a,Cristini2019a}. It is not clear which implementation for the overshoot yields the best results. It seems that different implementations should be used depending on the part of the star and the phase considered \citep{Viallet2015a}. In all these implementations, there are one or several free parameters needing a calibration. Depending on the choice of the calibrators domain and the type of constrain used (width of the MS band, velocity drop at the end of the MS), different values can be obtained \citep{Martinet2021a}.

In the interior layers, convection is usually treated as adiabatic, with an instantaneous mixing of chemicals and angular momentum. In advanced phases though, the nuclear timescale becomes shorter than the convective turnover timescale, and convection needs to be treated as a diffusion. In the outer layers, the adiabaticity is no more fulfilled, and the convection is usually treated in the MLT approximation. In this framework, convective cells of a fixed size are supposed to move through a fixed distance in the star, before being dissolved. This distance, the mixing length, is a function of the pressure scale height  $H_P$: $\ell=\alpha_\text{MLT} H_P$. $\alpha_\text{MLT}$ is a free parameter needing a calibration, which is usually the Sun. While after calibration, a fixed value is used for the whole range of masses and metallicity, there are indication that $\alpha_\text{MLT}$ should vary as a function of the mass and the {\teff} \citep{Pinheiro2013a}.

\subsection{Rotation}
Convective zones in the star drive a very efficient mixing of both the angular momentum and the chemical species. However, it has been shown that to reproduce the observed surface enrichment, mixing should also occur in the radiative zones of the star \citep{Maeder1987a,Langer1991a}. Rotation brings naturally such a mixing source. Everything in the Universe is in movement, often in rotation, and stars do obey this law of nature. Given the conservation of angular momentum through the orders of magnitude difference between a star-forming cloud and the star born from it, the question is not so much if stars are rotating but how come they are not rotating faster? In any case, rotation is an important ingredient in stellar modelling. It brings a support for gravity thanks to the centrifugal force, but also deforms the star since the centrifugal force is stronger at the equator than at the poles. This deformation creates a more or less pronounced gradient of temperature from the cooler equator to the hotter poles, driving a large-scale circulation \citep{Eddington1925a,Vogt1925a} known as the {\it meridional circulation}. It also breaks the spherical symmetry that is usually assumed in the modelling.

\subsubsection{Stellar structure with rotation}

 It has been shown \citep{Kippenhahn1970a} that the spherical stratification prevailing in non-rotating stars can be replaced by a rotationally-deformed stratification if the potential is conservative, which is the case for rotation. In such a case, the pressure, temperature, and density are constant on an equipotential surface $S_\Psi$ of volume $V_\Psi$. Each equipotential can be defined by a pseudo radius $r_\Psi$ that is the radius that would correspond to a sphere of volume $V_\Psi$: $V_\Psi = \frac{4\pi}{3}r_\Psi^3$. The Lagrangian coordinate $M_r$ becomes $M_\Psi$, the mass enclosed in the equipotential surface. For any quantity $f$ that is not constant on the equipotential surface, like it is the case with gravity $g$, we can take a mean value of the form $\left<f\right>=\frac{1}{S_\Psi}\int_{\Psi=\text{const.}} f d\sigma$, with $d\sigma$ an element of the surface with $\Psi=\text{const.}$.

When passing from the spherical definition to the equipotential definition, the 1D stellar structure equations are still valid, except that a correcting factor ($f_P$) must be applied to the hydrostatic equilibrium equation:
\begin{equation}
\frac{\partial P}{\partial M_\Psi} = -\frac{G M_\Psi}{4\pi r^4_\psi}\,f_P
\end{equation}
with $f_P=\frac{4\pi r^4_\Psi}{GM_\Psi S_\Psi}\frac{1}{\left<g^{-1}\right>}$. The radiative transfer equation also has to be modified and becomes:
\begin{equation}
\frac{\partial\ln T}{\partial M_\Psi}=-\frac{GM_\Psi}{4\pi r^4_\Psi}f_P\,\text{min}\left(\nabla_\text{ad},\nabla_\text{rad}\frac{f_T}{f_P}\right)
\label{Eq:dTdM}
\end{equation}
with $f_T=\left(\frac{4\pi r^2_\Psi}{S_\Psi}\right)^2\frac{1}{\left<g\right>\left<g^{-1}\right>}$.

For computing $f_P$ and $f_T$, the pseudo potential $\Psi = \frac{GM_\Psi}{r} + \frac{1}{2}\Omega^2r^2\sin^2\theta=\text{constant}$ has to be evaluated for every mass shell inside the star. To do so a non dimensional variable is introduced:
\begin{equation}
x=\left(\frac{\Omega^2}{GM_\Psi}\right)^{1/3}r
\end{equation}
which represents the radius expressed in units of the critical radius at the equator. The pseudo-potential $\Psi$ becomes $\Psi'=\frac{1}{x}+\frac{1}{2}x^2\left(1-\mu^2\right)$, with $\mu=\cos\theta$. $\Psi'$ is expressed in units of $\Psi_\text{crit}=\left(GM_\Psi\Omega\right)^{2/3}$, and lays between the two extreme cases of no rotation and critical rotation, where the centrifugal force overcomes the gravitation force of the star. By introducing the ratio $\xi\equiv x_\text{equa}/x_\text{pole}$ (with $\xi=1$ for no rotation and 1.5 for the critical rotation), $\Psi'$ becomes:
\begin{equation}
\Psi'=\frac{\xi}{\left(2(\xi-1)\right)^{1/3}}.
\end{equation}
Note that since $\Psi'\rightarrow +\inf$ when $\xi\rightarrow 1$, the full calculation is restricted for the case of rotation with $\xi\in]1;1.5]$.

It is now possible to compute all the needed quantities $V'_\Psi$, $S'_\Psi$, $x'_\Psi$, and define the correcting factors $f_P(x_\Psi')$ and $f_T(x_\Psi')$. This method has the advantage of having $f_P$ and $f_T$ depend only on the (predictable) deformation of the star, thus they can be computed once, and tabulated for being used throughout the computation.

\subsubsection{Transport of angular momentum in radiative zones}

Unfortunately, the internal transport of angular momentum in radiative zones is still escaping an accurate description. While asteroseismology has revolutionised our view on the internal transport in low-mass stars, intermediate-mass and massive stars are still scarcely studied with this technique and we are yet to build a statistically significant sample to constrain the angular-momentum transport processes.

Several implementations for the description of angular-momentum transport exist or co-exist in stellar evolution codes: the advecto-diffusive transport, which mildly couples the core and the envelope \citep{Zahn1992a}; the transport by magnetic fields, which yields a maximal core-envelope coupling \citep{Spruit1999a}; the transport by the internal gravity waves generated by convective plumes \citep{Talon2002a}. The different implementations available include one or several free parameters that need to be calibrated on observations.

The strength of the coupling between the core and the envelope is crucial as it affects the amount of angular momentum that can be stored in the core until the end of the evolution.

\subsection{Mixing of chemical elements in radiative zones}
The core-envelope coupling also influences the chemical stratification of the stars, since a mild coupling, yielding a strong gradient of angular momentum, would generate a shear between the layers rotating differentially, and hence a strong diffusion of chemical elements. The chemical stratification in turn influences the structure itself. The efficiency of the mixing is a multivariate function of the mass, metallicity, age, initial rotation rate, and multiplicity \citep{Maeder2009b,Georgy2013a}. The more massive the star, the more efficient is the mixing, so a surface enrichment will be present earlier on the main sequence that in the case of lower mass stars. The older the star, the more time it has had to be mixed, so the more enriched it is expected to be. At low metallicity, stars are more compact, so the mixing timescale is shorter: $\tau_\text{mix}\propto D/R^2$ with $D$ the diffusion coefficient and $R$ the size of the region to mix. When the mixing is extremely efficient, it can connect almost the full star, yielding what is called the {\it chemically quasi-homogeneous evolution} (CHE). This situation arises during the main sequence, mainly at low metallicity or in some cases of rapid rotation and very massive stars. Instead of expanding, the star remains extremely compact and blue \citep{Maeder1987b,Langer1992a}. This peculiar evolutive path is supposed to possibly give rise to long Gamma-ray bursts at the time of the supernova\footnote{see also the \encyclochap{Gamma Ray Bursts}}, the {\it collapsar} scenario, a phenomenon linked to the collapse of a massive hydrogen-free rapidly-rotating star \citep{Woosley1993a,Yoon2006a}.

Besides rotation, the internal gravity waves excited by the drumming of the convective plumes on the adjacent radiative layers have also been invoked to drive a mixing in the radiative zone \citep{Press1981a}. With all the intrinsic variations of the strength of the mixing, the choice of the calibrators is very delicate, requiring a good balance between a clean homogeneous sample and a size sufficiently large to be statistically significant. To transform observations into constrains is very complicated and requires consistent samples and a thorough analysis.

\section{Multiplicity}
On top of all the uncertainties presented above, both the mass and angular-momentum budget are strongly affected by interactions in gravitationally-bound systems\footnote{see also the \encyclochap{Evolution of binary stars}}. Massive stars have been shown to live preferentially in binary or multiple systems \citep{Mason2009a,Sana2013a,Dunstall2015a}. Interactions in such systems can change the mass budget: mass-transfer episodes can occur if one of the component fills its Roche lobe, $R_\star >= R_\text{Roche} = a \frac{0.49\,q^{2/3}}{0.6 q^{2/3} +\ln(1+q^{1/3})}$, with $a$ the separation, and $q=M_2/M_1$ \citep{Eggleton1983a}. The other component can accrete all or part of the mass lost. In some extreme cases, the accretor can become more massive than the donor, reversing $q$ (like in Algol systems).

Binary interactions also modify the angular-momentum budget and its internal repartition. Mass-transfer episodes transfer also angular momentum from one component to the other, spinning the accretor up. Tides dissipation exert a friction that adds a supplementary mixing of angular momentum and of chemicals, influencing the chemical stratification and the surface velocity.

As 70\% of massive stars have interacted or will interact during their life \citep{Sana2012a}, the observational constrains we could gather about stellar physics must be very carefully assessed. Wide binaries are interesting candidates as '{\it best single stars}' \citep{deMink2011a} but the large fraction of hierarchical multiple systems in the wide orbit "binaries" \citep{Tokovinin2008a} forces us to remain very cautious when using them for constraining stellar physics..

Massive binaries will end their life as NS+BH or BH+BH systems. In 2015, the first detection of gravitational waves arising from a BH+BH merger \citep{Abbott2016a} has opened a new window to study the end point of binaries. The discoveries done since then have posed a new challenge to stellar physicists: how to form such massive black holes and have them merge in a Hubble time\footnote{see also the \encyclochap{Population Synthesis of Gravitational Wave Sources}}? 

\section{Evolutionary sequence and nucleosynthesis\label{sec:evol_nuc}}
\subsection{Main sequence\label{sec:ms}}
Massive stars need a strong support for their mass and luminosity, so they are powered by the CNO cycle during H-burning, where the elements C, N, and O are catalysts in a chain of reactions producing a helium nucleus in the end:
$$\isotop{12}{C}(p,\gamma)\isotop{13}{N}(\beta^+)\isotop{13}{C}(p,\gamma)\isotop{14}{N}(p,\gamma)\isotop{15}{O}(\beta^+)\isotop{15}{N}(p,\alpha)\isotop{12}{C}.$$
In some cases, the isotope {\isotop{15}{N}} releases a photon instead of an $\alpha$, and the sub-cycle CNO II is initiated:
$$\isotop{15}{N}(p,\gamma)\isotop{16}{O}(p,\gamma)\isotop{17}{F}(\beta^+)\isotop{17}{O}(p,\alpha)\isotop{14}{N}.$$
In the hot cores of massive stars, there is a probability that the {\isotop{17}{O}} does not release an $\alpha$ but a photon, creating the CNO III branch:
$$\isotop{17}{O}(p,\gamma)\isotop{18}{F}(\beta^+)\isotop{18}{O}(p,\alpha)\isotop{15}{N}.$$
While the energy generation through $pp$-chains have a temperature dependence of $T^4$, that of the CNO cycle has a temperature dependence of $T^{17}$, that is way more reactive to any change in temperature. The large energy generation triggers convection in the core, so massive stars on the main sequence (MS) present a typical structure of a convective core surrounded by a radiative envelope. The convective core can occupy a large fraction of the star, reaching 60\% in mass fraction, or even more, in the upper mass end. Such large cores facilitate the appearance of nucleosynthetic products at the surface already during the MS.

\begin{figure}[t]
\centering
\includegraphics[width=.75\textwidth]{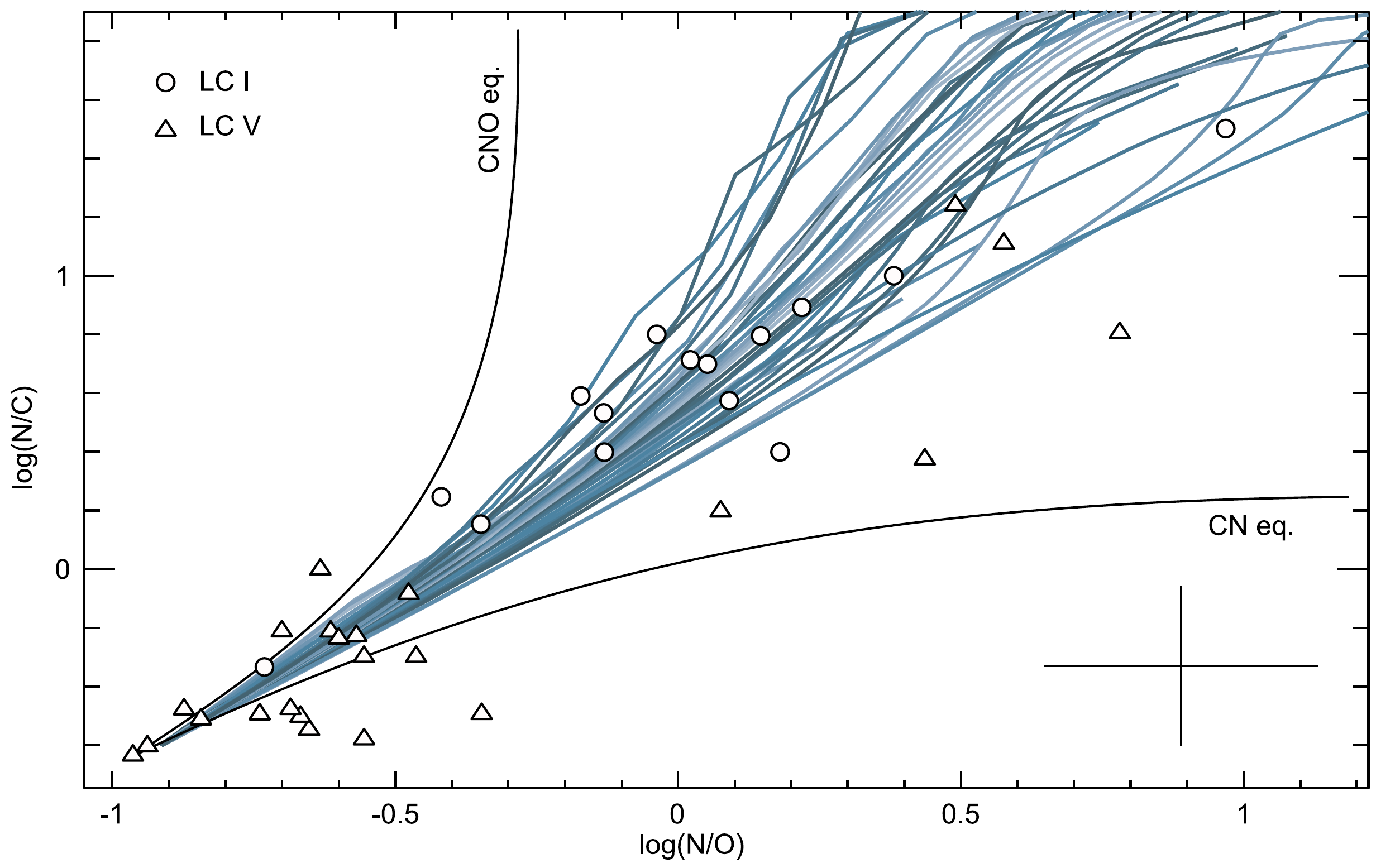}
\caption{N/C versus N/O (in numbers) diagram for the same models as Fig.~\ref{fig:mist_hrd}. The black lines show the two limiting assumptions of CN equilibrium or full CNO equilibrium. Observations of massive O stars with luminosity class V (triangles) and I (circles) from \citet{Martins2015a} are over-plotted. A typical error bar is indicated in the bottom right corner.}
\label{fig:mist_ncno}
\end{figure}

In the main CNO loop, the slowest reaction is $\isotop{14}{N}(p,\gamma)\isotop{15}{O}$. The signature of the active CNO cycle is hence a shift of carbon and oxygen to nitrogen. Note that the sum of species is constant, there is no increase in metallicity through the CNO cycle. The hotter the star, the quicker C is transformed into N, while O needs a larger temperature and the ON loop is less probable than the CN loop. Typically, massive stars have the CN cycle reaching equilibrium very quickly in the centre, with \isotop{12}{C} that can be considered as a constant. In this case, in a N/C versus N/O plane (C, N, and O in numbers), the evolution can be written as \citep{Maeder2014a}: $$\frac{\text{d}\log(\text{N/C})}{\text{d}\log(\text{N/O})}=\frac{1}{1+\text{N/O}}$$ If the temperature is lower, as in lower-mass stars or in more external regions in massive stars, it is the \isotop{16}{O} that can be considered as a constant, and the evolution in the N/C vs N/O plane can be written as: $$\frac{\text{d}\log(\text{N/C})}{\text{d}\log(\text{N/O})}=1+\text{N/C}$$ These two equations trace two limiting lines, the real abundances in stars laying somewhere in between (see Fig.~\ref{fig:mist_ncno}).

The core of a massive stars is hot enough to trigger other nucleosynthetic cycles like the NeNa or MgAl cycles. The NeNa cycle uses isotopes of neon and sodium as catalysts in a chain of proton captures and beta decays:
$$
\isotop{20}{Ne}(p,\gamma)\isotop{21}{Na}(\beta^+\nu)\isotop{21}{Ne}(p,\gamma)\isotop{22}{Na}(\beta^+\nu)\isotop{22}{Ne}(p,\gamma)\isotop{23}{Na}(p,\alpha)\isotop{20}{Ne}
$$
\noindent The {\isotop{23}{Na}} can also decay in the photon channel $\isotop{23}{Na}(p,\gamma)\isotop{24}{Mg}$, which gives rise to the MgAl cycle:
$$
\isotop{24}{Mg}(p,\gamma)\isotop{25}{Al}(\beta^+\nu)\isotop{25}{Mg}(p,\gamma)\isotop{26}{Al}(\beta^+\nu)\isotop{26}{Mg}(p,\gamma)\isotop{27}{Al}(p,\alpha)\isotop{24}{Mg}
$$
\noindent An important product of these cycles is {\isotop{26}{Al}}, a radioactive isotope whose characteristic 1.8 MeV emission is used to trace massive star formation in galaxies \citep{Diehl2004a}.

During the MS phase, the spectral types of massive stars range typically from early-B to O. The high surface temperature added to the high luminosity makes them powerful ionising sources in the Universe, and they are often surrounded by {\hii} regions. In that phase, their winds are thin and fast.

\subsection{Helium-burning phase\label{sec:heb}}
After H exhaustion, the core contracts and heats until it reaches the temperature needed to fuse He nuclei. Helium is fused through the triple-$\alpha$ reaction \citep[\tria][]{Salpeter1952a}, that takes place in two steps. First the fusion of two He nuclei creates a boron nucleus: $\isotop{4}{He}(\alpha,\gamma)\isotop{8}{B}$. This boron is unstable, with a lifetime of $6.7\times10^{-17}$\,s. However, there is a resonance level in the {\isotop{12}{C}} atom at 7.65 MeV, that corresponds to the energy of an {\isotop{8}{B}} + $\alpha$ compound, so this increases the probability of the boron nucleus to interact with a third $\alpha$ \citep{Hoyle1954a}. The {\tria} reaction is very reactive to temperature changes, with a dependence of the energy generation going as $T^{41}$. 

The {\tria} reaction builds {\isotop{12}{C}} in the core, but as the core continues heating up, the competing {\cago} reaction starts consuming the newly formed C nuclei. The more massive the star, the hotter the core becomes, and the stronger the {\cago} reaction becomes, depleting the core of its {\isotop{12}{C}}. This reaction is of prime importance in stellar nuclear physics, as it influences the advanced stages of the evolution (see Sect.~\ref{sec:advanced}), but unfortunately, it is still not known with the required accuracy \citep{deBoer2017a}.

Helium is also consumed through the reactions $\isotop{16}{O}(\alpha,\gamma)\isotop{20}{Ne}$ and $\isotop{20}{Ne}(\alpha,\gamma)\isotop{24}{Mg}$ but these reactions become significant only at $T>0.3$\,GK. The {\isotop{14}{N}} left from the previous H-burning episode reacts with the He nuclei in the reaction sequence $\isotop{14}{N}(\alpha,\gamma)\isotop{18}{F}(e^+\nu)\isotop{18}{O}$. The neutron-rich oxygen reacts then with an $\alpha$, creating a neutron-rich Ne through $\isotop{18}{O}(\alpha,\gamma)\isotop{22}{Ne}$. Ne in turn can react with an $\alpha$ through two channels: $\isotop{22}{Ne}(\alpha,\gamma)\isotop{26}{Mg}$ or $\isotop{22}{Ne}(\alpha,n)\isotop{25}{Mg}$. The last reaction is important as it releases a neutron which feeds the so-called 'weak $s$-process' nucleosynthesis (slow neutron capture) in massive stars, a process that is at the origin of elements with $A\simeq88$, but also of lighter elements such as \isotop{36}{S}, \isotop{37}{Cl}, \isotop{40}{Ar}, \isotop{40}{K}, and \isotop{45}{Sc}.

Just after central H exhaustion and during He burning, massive stars divide into two separate behaviours \citep{Conti1975a}. The least massive ones expand rapidly after central H exhaustion, and become Red Supergiants (RSG). The most massive ones remain in the blue part of the HRD, and may become Wolf-Rayet stars (WR) eventually. In a narrow mass range between the two, stars evolve to RSG after central H-exhaustion and then go back towards the blue part of the HRD to become yellow supergiants (YSG), LBV, or even WR. The transition mass depends on  the physics considered, but lies roughly between 30-40 \msol. Given the short lifetime of evolved stars of that mass range, clusters hosting both RSG and WR stars are not expected to be very numerous, except if the WR stars are a result of binary interactions \citep{Meynet2015a}.

\subsection{To the red supergiant stage\label{sec:rsg}}
The H-burning shell that builds at the end of the central H-fusion phase acts like a mirror for the structure of the star: the core contraction translates into a rapid envelope expansion, cooling down the surface until an outer convective zone appears and deepens inside the envelope. The rapid mixing inside the outer convective zone dredges some nucleosynthetic products up to the surface, increasing even more the opacity. When the surface has cooled down to $\log(\teff)<3.7-3.6$, the star settles on the RSG branch\footnote{see also the \encyclochap{Giants, supergiants and hypergiants}}. In this phase, the luminosity is essentially set by the mass of the He-burning core \citep{Farrell2020a}, which grows through time.

The driving mechanism of the wind changes. It is not yet fully clear what is the exact mechanism at play: pulsation, dust production, a combination of both, or something else \citep{Levesque2017a}. It is not clear either whether it is a steady wind, eruptive episodes, or a combination of both. In any case, the link between the winds and the chemical composition of the star becomes less strong than it is the case for radiatively-driven winds. RSG winds are thick and slow and leave a lasting imprint in the circumstellar medium, as highlighted by early SN observations \citep{Chevalier2006a,Moriya2021a}. Although RSGs might lose a lot of mass in this phase, the core mass is not affected by the change in the total mass, as shown by \citet{Farrell2020b}: the H-burning shell acts as a buffer and adapts its power to the mass of the envelope it has to support, leaving the core untouched. However, if the mass lost is large, and the core-mass to total-mass ratio increases above 0.6, the star leaves the RSG branch and evolves back to the blue.

In the lower end of the massive stars mass domain, RSG might undergo a blueward loop that will make them cross the Cepheid instability strip. While in the intermediate-mass domain, the appearance of the Cepheid loops is progressive in more and more massive stars, the end of the loop behaviour in the massive stars domain is abrupt, and seems to be related to a limiting luminosity \citep{Anderson2014a}. In models, the loop behaviour is extremely sensitive to the physical ingredients that have an influence on the size of the core (like overshooting) or the evolution of the mass (like winds).

Stars exploding while in the RSG phase are supposed to create Type II supernovae. Almost all Type II SN arise from progenitors with a luminosity $\log L/L_\odot\lesssim 5.1$, pointing to a maximal mass of $\sim$18\,{\msol} above which the stars directly collapse into black holes or have moved back to the blue and explode as Type Ib or Ic supernovae \citep{Smartt2015a}. 

\subsection{To the Wolf-Rayet stage\label{sec:wr}}
For the high-mass end of massive stars, the strong winds during MS peel off the outer layers down to the burning regions, which let them enter the WR stage either still on the MS or early in the He-burning phase\footnote{see also the \encyclochap{Wolf Rayet stars}}. After the hydrogen-rich envelope is gone, the first layers to be uncovered are those that have undergone the CNO H burning and that are rich in nitrogen. When the winds have removed the H-burning region, the He-burning core is exposed, revealing C- and O-rich layers \citep{Crowther2007a}.

WR winds are thick and fast, and the photosphere is moved outwards inside the wind, shifting the estimation of the {\teff} to a lower value. The clumpy structure of the wind could be inherited from sub-surface convective zones driven by the iron opacity \citep{Cantiello2009a}.

The mass loss of previous stages (MS or RSG) is of prime importance for the formation of WR stars out of single stars. The minimum mass for a single star to become a WR is thus metallicity dependent \citep{Shenar2019a}, higher at lower metallicity. Lower-mass WR are produced through mass exchange in binary systems.

\begin{figure}[t]
\centering
\includegraphics[width=.5\textwidth]{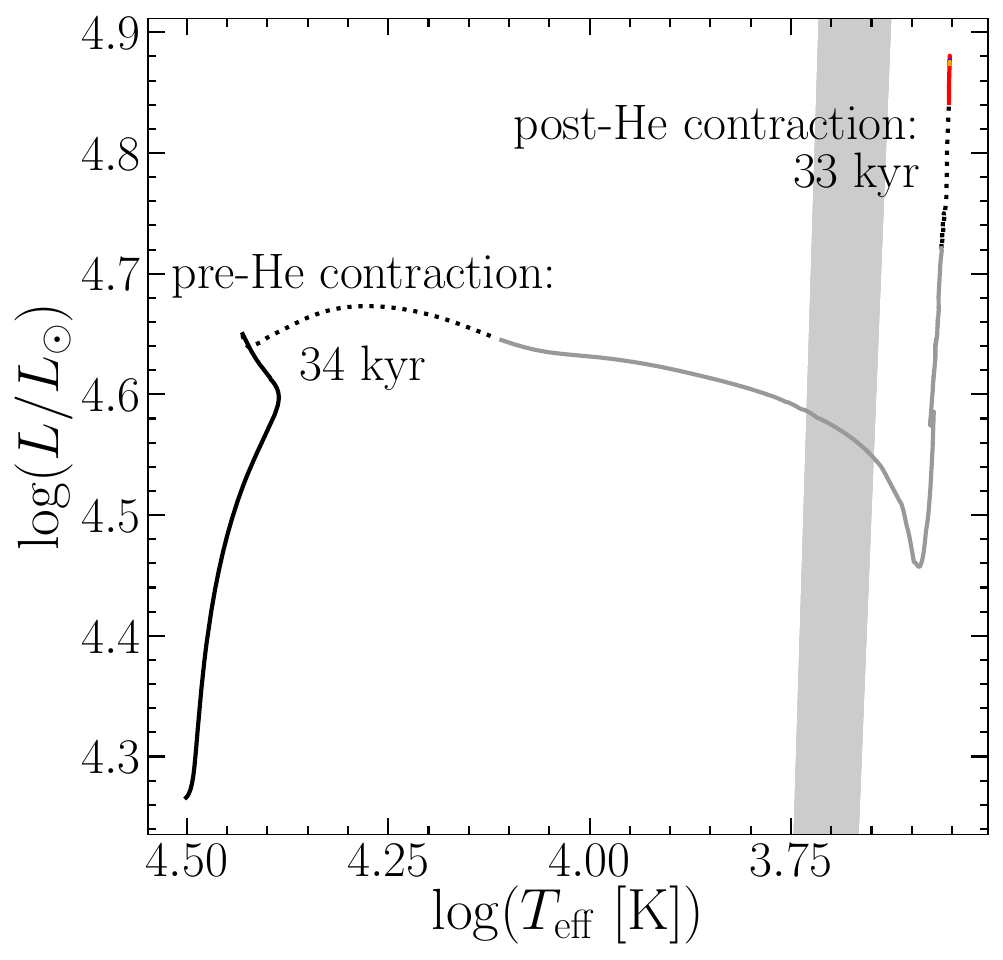}\includegraphics[width=.5\textwidth]{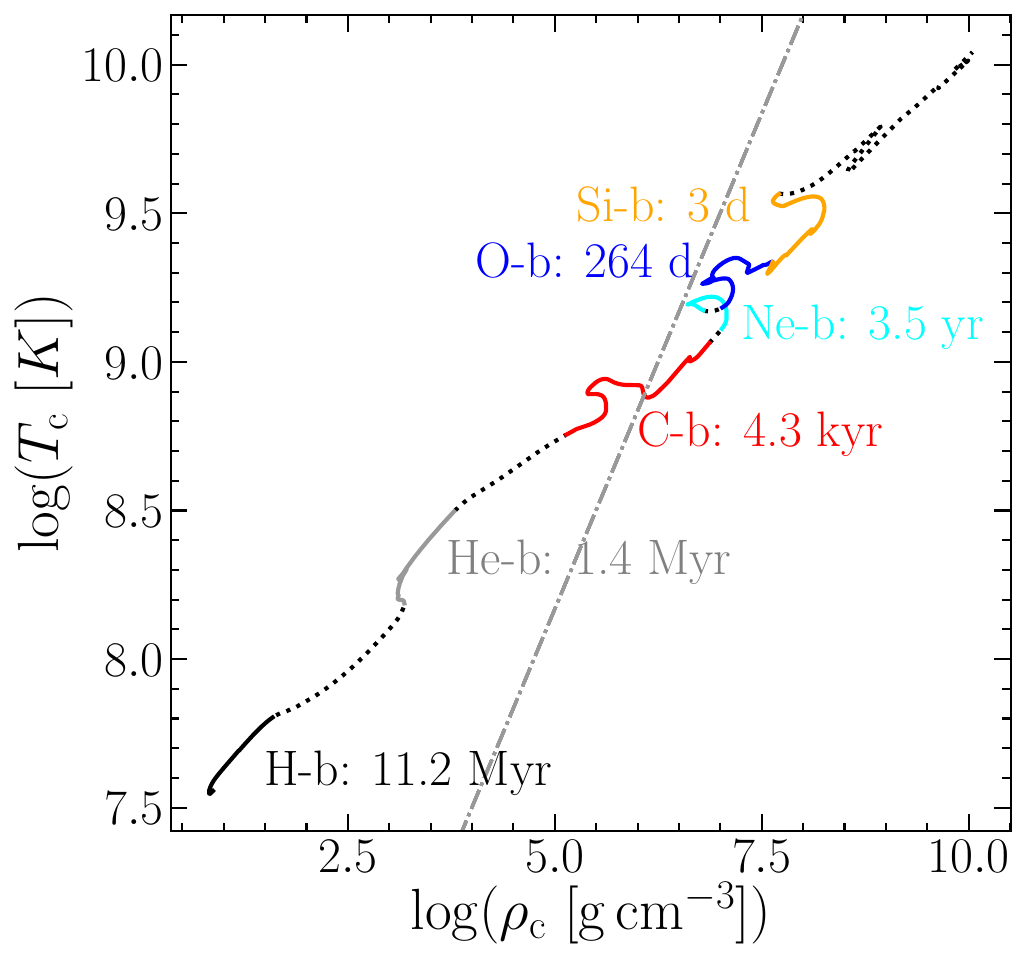}
\caption{Evolution of a 15\,{\msol} model at solar metallicity through the advanced phases \citep[model from][]{Griffiths2025a}. {\it Left:} Surface conditions in a Hertzsprung-Russell diagram. The shaded zone is the Cepheid instability strip. The time span of the contraction phases (black dotted lines) before and after central He-burning is indicated.; {\it Right:} Central conditions in a $\log(T_\text{c})$ vs $\log(\rho_\text{c})$ diagram. The dot-dashed line shows the limit between non-degenerate and degenerate gas. The time span of the burning phases is indicated.}
\label{fig:advanced}
\end{figure}
\subsection{Advanced phases\label{sec:advanced}}
While the H- and He-burning phases behaviour depends on the initial mass and chemical composition of the star, the advanced phases are independent of these parameters but strongly dependent on two new ones: the CO-core mass fraction ($M_\text{CO}$) and the abundance of {\isotop{12}{C}} left at the end of He-burning \citep{Chieffi2020a}. $M_\text{CO}$ is usually defined as the mass coordinate where the abundance of C+O is larger than 0.75 in mass fraction, or alternatively the mass coordinate below which the He abundance drops below $10^{-3}$. In any case, it marks a sharp transition between the region of the star that has been fully processed by He burning, and the region above.

After He burning, the internal conditions become favourable for neutrino production \citep[out of $e^\pm$ pair annihilation, plasmon decay, or photo-production, see][]{Beaudet1967a}, leading to an energy loss for the star, roughy proportional to $T^9$. The energy leak for the star as well as the ever heavier fuels which require a high temperature to burn lead to a strong acceleration of the evolution \citep{Woosley2002a}. While the core evolves on timescales which become shorter and shorter, the envelope evolves on a Kelvin-Helmholtz time $\left(t_\text{KH}=\frac{GM^2}{RL}\right)$ that is way longer. This leads to a decoupling between the core and the envelope after the central He exhaustion. Fig.~\ref{fig:advanced} shows the evolution of a 15\,{\msol} model in the traditional HRD, which presents the evolution of the surface conditions, and in the $T-\rho$ diagram, which presents the evolution of the central conditions.
From the C-burning phase onwards, the surface conditions are unable to transcribe the core's evolution, even though its temperature changes by more than 1 order of magnitude, and its density by almost 5 orders of magnitude.

\subsubsection{Carbon fusion}
\begin{figure}[t]
\centering
\includegraphics[width=.85\textwidth]{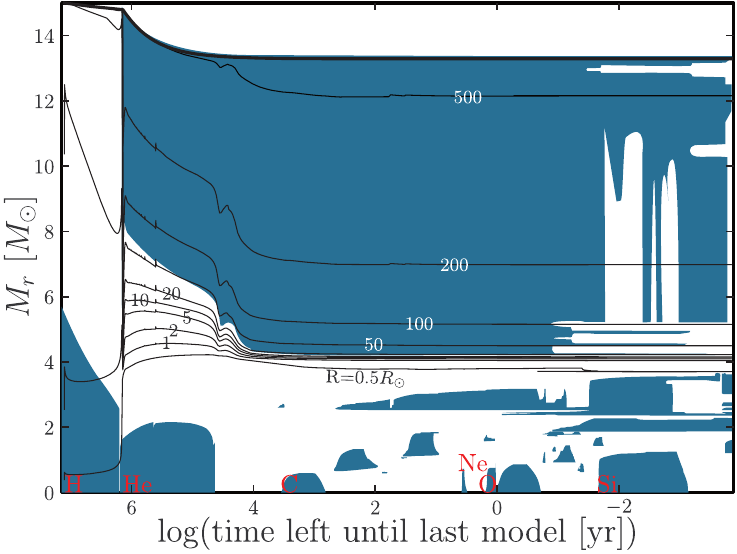}\hspace{1em}\includegraphics[width=.85\textwidth]{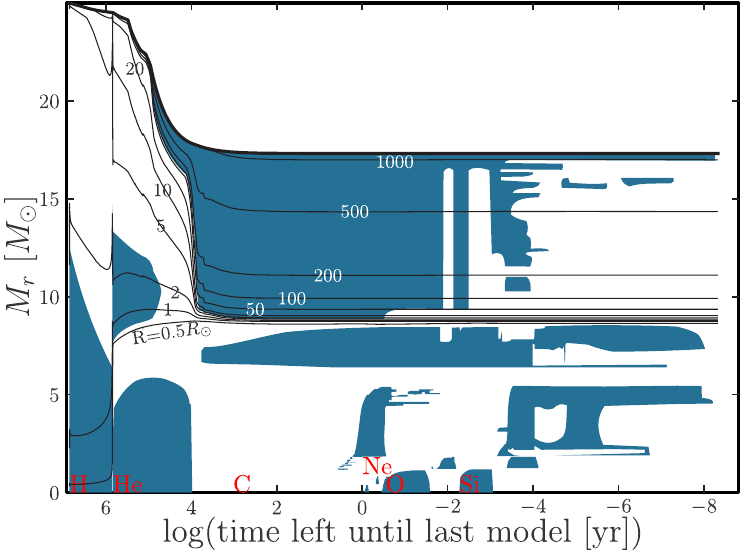}
\caption{Kippenhahn diagrams of a 15\,{\msol} (top) and a 25\,{\msol} (bottom) models, showing the evolution of the properties of the mass coordinates inside the star (surface is up, centre is down) along a time axis that is $\log(t_\text{fin}-t)$. Convective zones are shaded in blue. The burning phases are indicated in red. Lines of iso-radii are drawn. Models from \cite{Griffiths2025a}.}
\label{fig:kipp}
\end{figure}
The fusion of two {\isotop{12}{C}} yields a highly excited {\isotop{24}{Mg}} nucleus with an excess energy of about 14 MeV. The most probable channels hence imply the emission of light particles rather than a photon: $\isotop{12}{C}(\isotop{12}{C},\alpha)\isotop{20}{Ne}$, $\isotop{12}{C}(\isotop{12}{C},p)\isotop{23}{Na}$, or $\isotop{12}{C}(\isotop{12}{C},n)\isotop{23}{Mg}$. The $\alpha$ and $p$ branches have close to the same probability, while the $n$ branch remains very small even at high temperature. The temperature dependence of the energy generation by the full {\isotop{12}{C}} fusion is $T^{28}$.

The complex chemical composition left by He burning and weak $s$-process added to the different light particles emitted by C-burning and the various daughter elements makes the full nucleosynthetic landscape very complicated, with hundreds of isotopes involved. The main nucleosynthetic products of the C-burning phase are {\isotop{16}{O}}, {\isotop{20}{Ne}}, and {\isotop{23}{Na}}, with also neutron-rich isotopes of neon and magnesium.

Depending on the mass of the star, C-burning takes place convectively or radiatively. Figure~\ref{fig:kipp} shows the internal structure of a 15\,{\msol} model (top) and a 25\,{\msol} model (bottom), the 15\,{\msol} showing a convective C burning, while the 25\,{\msol} showing a radiative C burning. The transition from convective to radiative central C burning occurs actually for CO cores with a C mass fraction $X_\text{C}\lesssim 0.25$, so for an initial mass of around 20-22\,{\msol}, the more massive star losing too much energy through neutrino emission for the nuclear reactions to trigger convection \citep{Schneider2021a}. In stars with a mass below $\sim$22\,\msol, after central C exhaustion, a series of 2-3 subsequent convective shells appear before the core is massive enough for being able to contract and heat up to the temperature of Ne disintegration. The succession and strength of C-burning convective shells depend on the fraction of {\isotop{12}{C}} left after He burning, and play a crucial role in the internal structure of the star by influencing the compactness of the star. Stars more massive than 22\,{\msol} have a strong contraction after C exhaustion, which makes the C-burning shell to be very powerful and largely convective, with the result of reducing the compactness of the layers above \citep{Laplace2024a}.

\subsubsection{Neon disintegration}
Although {\isotop{16}{O}} is more abundant and 'next in line' in terms of Coulomb barrier, its double magic nature (8 protons + 8 neutrons) makes it more stable than neon, that becomes the next fuel when the temperature reaches 1.5 GK during the post C-burning contraction. Neon is not fused in stars, it is rather photo-disintegrated through the reaction $\isotop{20}{Ne}(\gamma,\alpha)\isotop{16}{O}$. The energy generation by the Ne photo-disintegration is very sensitive to temperature, with $T^{50}$.

The {\isotop{16}{O}} created by Ne disintegration can react back with the $\alpha$ released to directly form the {\isotop{20}{Ne}} again, but after the reactions come to equilibrium, the abundance of {\isotop{16}{O}} steadily increases. Ne burning liberates $\alpha$ particles, and $\alpha$ captures build up isotopes up to silicon: $\isotop{20}{Ne}(\alpha,\gamma)\isotop{24}{Mg}$, and $\isotop{24}{Mg}(\alpha,\gamma)\isotop{28}{Si}$. Side reactions involving Mg isotopes and neutron channels build a complex network of isotopes of Mg, Al, Si, and P. At the end of Ne burning, the core is mainly composed of {\isotop{16}{O}}, {\isotop{24}{Mg}}, and {\isotop{28}{Si}}, with some neutron-rich isotopes of magnesium and silicon, as well as aluminium isotopes and {\isotop{31}{P}} and {\isotop{32}{S}}.

In order for a star to proceed into Ne burning, it has to have a CO-core mass larger than 1.37\,{\msol} \citep{Nomoto1984b}. Stars with $M_\text{CO}$ only slightly more massive experience a Ne ignition that is off-centre in degenerate conditions, and the flame proceeds inwards to the centre by steps of burning episodes followed by contraction episodes. In this case, the density of the core can increase beyond the critical density for electron captures on {\isotop{24}{Mg}} to occur. Since the degenerate matter is sustained by the electrons, these captures remove a source of pressure and lead to a kind of supernova that is called electron-capture supernova \citep[EC-SN,][]{Barkat1974a,Miyaji1980a}, occurring before an iron core is formed (during Ne- or O- burning). The exact mass range of stars evolving to EC-SN is very dependent on the physics considered \citep[rotation, mass loss, CBM,][]{Siess2007a,Jones2014a} but is around 8-10\,{\msol}.

\subsubsection{Oxygen fusion}
When the central temperature reaches $2\cdot10^9$~K, oxygen is able to fuse. Like two {\isotop{12}{C}} yielding an excited state of {\isotop{24}{Mg}}, the fusion of oxygen yields an excited state of {\isotop{32}{S}} whose excess energy (16.5 MeV) is efficiently evacuated by the release of a light particle: $\isotop{16}{O}(\isotop{16}{O},\alpha)\isotop{28}{Si}$, $\isotop{16}{O}(\isotop{16}{O},p)\isotop{31}{P}$, or $\isotop{16}{O}(\isotop{16}{O},n)\isotop{31}{S}$. The probability of the $\alpha$ channel is 2/3 that of the proton channel. The neutron channel has a probability of less than 10\%. If the temperature is very high, a deuterium can be produced: $\isotop{16}{O}(\isotop{16}{O},\text{D})\isotop{30}{P}$. The energy generation by O burning is less dependent on temperature than Ne, going as $T^{34}$.

At the end of central O-burning, the core is composed of 90\% of {\isotop{28}{Si}} and {\isotop{32}{S}}. The rest is composed by all sorts of side reactions yielding isotopes of Cl, Ar, K, and Ca mainly. Weak interactions become non negligible at that stage, and a clear neutron excess appears. Given the high temperature of O fusion, the $s$-process elements built in previous stages start melting down to Fe through photo-disintegration, and disappear from the core.

In the high-mass end of the massive stars domain, and in particular at low metallicity, oxygen burning is the stage at which the central conditions become favourable to the creations of electron-positron pairs. The subsequent loss of energy and pressure destabilises the star by lowering the adiabatic gradient $\Gamma1=\frac{\partial P}{\partial \rho}$ below the stability limit of 4/3. To enter this regime, stars need to have a large He core mass \citep{Woosley1986a}. For $M_\text{He}\gtrsim 40\,\msol$, pair instability drives pulsations, inducing a heavy mass loss and reducing the mass of the remnant BH, while in the case of $M_\text{He}\gtrsim 60\,\msol$, the first pulse is energetic enough to trigger a powerful explosion leaving no remnant at all \citep{Belczynski2016a}. The link between the He-core mass and the initial stellar mass depends on the physics considered, but also strongly on the metallicity considered: since mass loss is less efficient at low metallicity, the stars retain most of their mass and build a larger core to sustain it.

\subsubsection{Silicon burning, nuclear statistical equilibrium and pre-collapse}
As in the case of neon, silicon burning is not a fusion but a photo-disintegration, or actually a sequence of photo-disintegrations. The main channel is the $\alpha$ channel $\isotop{28}{Si}(\gamma,\alpha)\isotop{24}{Mg}(\gamma,\alpha)\isotop{20}{Ne}(\gamma,\alpha)\isotop{16}{O}(\gamma,\alpha)\isotop{12}{C}(\gamma,2\alpha)\alpha$. In parallel, the many $\alpha$ produced interact back with the nuclei from any step of the chain, but in particular from {\isotop{24}{Mg}} and {\isotop{28}{Si}}. The combination of photo-disintegration and $\alpha$ capture builds up clusters of isotopes in quasi statistical equilibrium (QSE), gathered around the tightly bound Fe-group elements.

Progressively with the declining {\isotop{28}{Si}}, all the forward reactions become at equilibrium with their reverse, and all the species settle on QSE. Only the weak interactions cannot reach equilibrium, since the neutrinos produced escape the star immediately. Given the very high temperature and density, the favoured isotopes are not the most tightly bound, but those that are close to the mean neutron excess $\eta=\sum_i (N_i-Z_i)Y_i$ (where $N_i$ is the number of neutrons, $Z_i$ the number of protons, and $Y_i$ the relative abundance of the isotope $i$ in number). At the end of core Si burning, weak interactions have increased the neutronisation of the stellar matter and $\eta$ is of the order of 0.01 -- 0.1 \citep{Chieffi1998a}.

Once the core is composed of Fe-peak elements only, it cannot extract any nuclear energy anymore. The Si shell burning adds to the mass of the iron core until it becomes unstable and starts collapsing\footnote{see also the \encyclochap{Core collapse supernovae}}. The number and locations of these Si shells vary a lot depending on the conditions for shell C burning, and on the location and strength of the O-burning shell. Figure~\ref{fig:kipp} (bottom) shows how the Ne/O/Si shells can merge, creating large convective zones. The mass at which the Fe core starts to collapse is the Chandrasekhar mass, but it needs to be corrected by several factors, including the electron density, the entropy, and the boundary pressure of the outer layers compressing the core \citep{Woosley2002a}.

\section{Summary}
By going through all the stages of nuclear burning, massive stars are important providers of new heavy chemical elements. They are key contributors to the chemical evolution of the Universe. By injecting kinetic energy in the interstellar medium, they influence star formation. Their strong radiation dominate the spectra of galaxies.

Although they are crucial actors in the evolution of galaxies, massive stars are still very mysterious. While the nucleosynthetic path through all the phases of nuclear burning is well known, many processes playing a key role in their evolution are still escaping an accurate description \citep{Laplace2024a}. 
\begin{ack}[Acknowledgments]
_The author is acknowledgeable to the Swiss National Fund project 212143 and to the STAREX grant from the ERC Horizon 2020 research and innovation programme (grant agreement No. 833925).
\end{ack}

\seealso{\null\\
\citet{Woosley2002a}\\\citet{Langer2012a}\\\citet{Iliadis2007a}}

\bibliographystyle{Harvard}
\bibliography{BibTexRefs}

\end{document}